\newcommand{\tiu}{\joule~\metre^{-2}~\second^{-0.5}~\kelvin^{-1}}

\newcommand{\F}{(175706) 1996 FG$_{3}$}
\newcommand{\FS}{1996 FG$_{3}$}
\newcommand{\pv}{\ensuremath{p_V}}

\newcommand{\km}{\kilo\metre}
\newcommand{\Fernandez}{Fern\'{a}ndez}

\documentclass[12pt,preprint]{aastex}

\usepackage{xcolor}

\usepackage[squaren,textstyle]{SIunits}
\usepackage{url}

\usepackage{amssymb}

\begin{document}

 \title{Physical characterization and origin of binary
    near-Earth asteroid (175706) 1996 FG$_{3}$
\footnote{Partially based on observations
    obtained at 
    the European Southern Observatory (ESO, program ID 383.C-0179A).
    Observations were also obtained at the Infrared
    Telescope Facility, which is operated by the University of Hawaii
    under Cooperative Agreement no.\  NCC 5-538 with the National
    Aeronautics and Space Administration, Science Mission Directorate,
    Planetary Astronomy Program.}}

\author{Kevin J. Walsh}
\affil{Southwest Research Institute, 1050 Walnut St. Suite 400, Boulder, CO, 80302, USA}
\email{kwalsh@boulder.swri.edu}

\author{Marco Delbo'}
\affil{UNS-CNRS-Observatoire de la C\^ote d'Azur,
BP 4229, 06304 Nice cedex 04, France}

\author{Michael Mueller}
\affil{UNS-CNRS-Observatoire de la C\^ote d'Azur,
BP 4229, 06304 Nice cedex 04, France \\
Low Energy Astrophysics, SRON, Postbox 800, 9700AV Groningen, Netherlands}

\author{Richard P. Binzel}
\affil{
Department of Earth, Atmospheric, and Planetary Sciences, Massachusetts Institute of Technology, Cambridge, Massachusetts 02139, USA
}

\author{Francesca E. DeMeo}
\affil{
Department of Earth, Atmospheric, and Planetary Sciences, Massachusetts Institute of Technology, Cambridge, Massachusetts 02139, USA
}


\begin{abstract}
  The near-Earth asteroid (NEA) (175706) 1996 FG$_{3}$ is a
  particularly interesting spacecraft target: a binary asteroid with
  a low-$\Delta v$ heliocentric orbit.
  The orbit of its satellite has provided valuable information about
  its mass density while its albedo and colors suggest it is primitive {
  or part of the C-complex taxonomic grouping}. We extend the physical
  characterization of this object with new observations of its
  emission at mid-Infrared (IR) wavelengths 
{ and with near-IR reflection spectroscopy.}
  We derive 
{ an area-equivalent system}
diameter of 
$1.90 \pm \unit{0.28}{\km}$ 
{ (corresponding to approximate component diameters of \unit{1.83}{\km} and \unit{0.51}{\km}, respectively)}
and a geometric albedo of $0.039 \pm 0.012$. 
\F\ was previously classified as a C-type asteroid, though the
combined 0.4--2.5 \micron\ spectrum { with thermal correction}
indicates classification as B-type; { both are consistent with the
  low measured albedo.}  Dynamical studies show that \F\ has most
probably originated in the inner main asteroid belt.  { Recent work
  has suggested the inner Main Belt (142) Polana family as the possible
  origin of another low-$\Delta v$ B-type NEA, (101955) 1999
  RQ$_{36}$.  A similar origin for \F\ would require delivery by the
  overlapping Jupiter 7:2 and Mars 5:9 mean motion resonances rather
  than the $\nu_6$, and we find this to be a low probability, but
  possible, origin.  }

\end{abstract}


\section{Introduction}
Among the $\sim$ 37 known NEAs with satellites, \F\ has a particularly
low $\Delta v$ value, making it an ideal target for a spacecraft
mission \citep{Perozzi2001,Christou2003,Johnston2010}. In fact, it is
the baseline target of the ESA mission study { Marco~Polo-R}
\citep{Barucci2011}.  \F , hereafter \FS , is a prototype for the
``asynchronous'' NEA binaries.  About 15\% of all NEAs and small main
belt asteroids (D $<$ 10 km) are estimated to be such a binary
\citep{Pravec2006}, and are characterized by a rapidly rotating
primary (P $<$ 4 h), a nearly spherical or oblate primary, and a
moderately-sized secondary on a close orbit \citep{Pravec2006}.

\FS\ was the first binary NEA for which eclipse events were detected in
optical wavelengths \citep{Pravec2000,Mottola2000}.  
{ From these and later observations, \citet{Scheirich2009} found the}
period of the mutual orbit to be $16.14\pm0.01$~hr, and a diameter
ratio of $D_2/D_1=0.28^{+0.01}_{-0.02}$.
A circular orbit is consistent with the data, but the preferred
solution has $e=0.10^{+0.12}_{-0.10}$ and semimajor axis $a$ $\sim3.4$
primary radii. The primary is an oblate ellipsoid with an axis ratio
around $a/b \sim$1.2, while the secondary is prolate with an axis
ratio around 1.4. The spin period of the primary is 3.6~hr, that of
the secondary is unknown. \citeauthor{Scheirich2009}\ also find a mass
density of \unit{$1.4^{+1.5}_{-0.6}$}{\gram\per\centi\metre\cubed}
and an ecliptic latitude of the mutual spin axis of  \unit{-84}{\degree}.



The formation of these binaries { points} to a history of spin rate
increase due to the thermal YORP effect, followed by re-shaping and
mass-loss ending with a satellite in a close orbit
(\citet{Walsh2008}; { see also \citet{Scheeres2007} work on fission of contact binary asteroids}). This process demands a ``rubble pile'' or
gravitational aggregate makeup, where bulk re-shaping is the cause for
the ubiquitous oblate ``top-shape'' and equatorial ridge \citep[see
  1999 KW$_4$,][]{Ostro2006}. The process may end with preferential
material migration from the poles of the primary towards the equator
and also with general regolith depletion \citep{Walsh2008,Delbo2011}.


Photometric observations determined the optical magnitude
$H=17.76\pm0.03$ and phase coefficient $G=-0.07\pm0.02$
\citep{Pravec2006}.  Based on thermal observations (3.6 and
\unit{4.5}{\micron}) using the ``Warm Spitzer'' space telescope,
\citet{Mueller2011} find an area-equivalent diameter of this binary
system of $1.8^{+0.6}_{-0.5}\ \kilo\metre$ and a geometric albedo of
$\pv=0.04^{+0.04}_{-0.02}$, consistent with a taxonomic classification
in the C complex.  It should be noted that their diameter and albedo
values are based on an assumed color temperature (as opposed to a
measured one), subjecting their results to significant systematic
uncertainty.  { \citet{Wolters2011}, observing at VLT, determined an
  effective diameter of 1.71$\pm$0.07~km and a geometric albedo of
  $\pv=0.044\pm0.004$; furthermore they estimate the thermal inertia 
  to be $\Gamma = 120 \pm \unit{50}{\tiu}$.}
\citeauthor{Mueller2011}\ also studied the thermal history of
\FS\ over its chaotic orbital evolution
finding that it was likely (90\%) heated above \unit{482}{\kelvin},
possibly hot enough to induce thermal alteration of previously
{ primitive surface material}.

\citet{Pravec2000} and \citet{Mottola2000} estimated that \FS\ is a
C-type body from its broadband color indices and the steep phase
dependence.  The second phase of the SMASS survey's visible wavelength
spectroscopy (0.44--0.92 micron) classified \FS\ as a C-type
{ \citep{Bus99}}, and when the spectroscopy of \FS\ was extended to 0.4
-- 1.6 $\mu$m by \citet{Binzel2001} \FS\ remained classified as a
C-type.  Recently, the taxonomic classification scheme of
\citet{Demeo2009} has leveraged data reaching 2.45 $\mu$m, though
\citeauthor{Demeo2009}\ did not initially examine \FS.
\citet{DeLeon2011} recently analyzed three spectra of \FS, finding
multiple acceptable taxonomic classifications; Ch, C, Xk and B-type.

In this study, we present new multi-wavelength observations,
reflectance spectroscopy { between 0.4 and 2.5} \micron, 
and thermal-infrared photometry from 8--\unit{19}{\micron} to extend the physical
characterization of \FS. We utilize dynamical modeling of main belt
source regions for NEAs and { analyze the possibility of the B-type
  Polana family as a source for \FS. }

\section{Thermal-IR observations and thermal modeling}

{ We measured the thermal emission of \FS\ from observations at
  mid-IR wavelengths (8--18.5 \micron) and also obtained its visible
  reflected light. Combined with a suitable model of the surface
  thermal emission we determined its size and albedo; results are
  given in Sect.~\ref{sect:neatm}}.

\subsection{IRTF-MIRSI observations}
\label{sect:mirsi}

We observed \FS\ on 1 May 2009 UT using the Mid-Infrared Spectrometer
and Imager MIRSI \citep{mirsi1,mirsi2} on the NASA Infrared Telescope
Facility (IRTF) on Mauna Kea, Hawaii.  We used photometric filters
centered at 8.7, 9.8, 11.6, and \unit{18.4}{\micron} with spectral
bandwidths just short of $\sim10\%$.  A standard four-point chop-nod
pattern was used to remove background flux.  The night was clear with
no perceptible clouds, { and the} atmospheric humidity decreased
from $\sim30\%$ to $\sim6\%$.  The \citet{CohenIV} flux calibrator
$\alpha$~Hya was observed before and after the target
at very similar airmass.  A second \citet{CohenIV} standard,
$\alpha$~Boo, was observed afterwards for cross-checks on the flux
calibration.  
The IRTF IDL library%
\footnote{\url{http://irtfweb.ifa.hawaii.edu/~elv/mirsi_steps.txt}}
was used to coadd data as needed and to remove known instrument
artifacts.  Raw signal counts were derived using standard synthetic
aperture procedures.  
Flux calibration factors and atmospheric extinction coefficients for
each filter were derived from the observations of $\alpha$~Hya.
Analyzing the calibrator observations (including $\alpha$~Boo) we
reproduce their \citet{CohenIV} fluxes to within 1\% for N-band
fluxes, and to within 4\% for Q-band (18.4 \micron).
Color corrections to asteroid fluxes are expected to be at the percent
level and were not applied.  Final MIRSI fluxes are given in
Table~\ref{tab:mirsi}.

In the N-band filters, we typically performed one or two back-to-back
repeat observations to ascertain repeatability,
which was within 10\%.  No such variation is seen in repeat
observations of the flux calibrators, likely due to fluctuations in
the background level.
Our observations cover a fraction of the 3.6~hr rotation period (with
a low-amplitude lightcurve), and given the reported repeatability, no
lightcurve correction was attempted.


%

MIRSI observations were interspersed with $V$-band photometry
observations using the optical CCD \emph{Apogee}, which was
calibrated against observations of \citet{Landolt} standard SA~101-57.
Due to the proximity in airmass ($<$0.05), extinction correction is
not critical; we assumed an extinction coefficient of 0.12
mag/airmass.  The 12 measured $V$ magnitudes are constant within the
photometric uncertainty of typically 0.15~mag per data point, no
indication of a $V$-band lightcurve was detected.  The average $V$ mag
measured is 16.94,
in excellent agreement with the expected $V$ magnitude of $16.94\pm0.13$ for lightcurve average \citep[based upon $H=17.76\pm0.03$,
$G=-0.07\pm0.02$,][]{Pravec2006}.

\subsection{VLT-VISIR observations}
\label{sect:visir}

Independent thermal-IR observations of \FS\ were obtained on 2 May 2009 UT
using the VLT Imager and Spectrometer for mid Infrared { \citep[VISIR;][]{Lagage2004}}
installed at the 8.2m VLT Melipal telescope of the European Southern
Observatory (ESO), Cerro Paranal, Chile.
Photometric observations were carried out through narrow-band filters
centered at 8.59, 11.88 and \unit{18.72}{\micron}. 
The observation design was largely analogous to that of our MIRSI
observations (see Sect.\ \ref{sect:mirsi}).  The
\citet{CohenIV,CohenX} standard star HD93813 was observed at an
airmass very similar to that of the science target and used for
absolute flux calibration.\footnote{see
  \url{http://www.eso.org/sci/facilities/paranal/instruments/visir/tools/zerop_cohen_Jy.txt}}
Data were reduced as described above, except in the case of the PAH\_2
filter centered at \unit{11.88}{\micron}, where coadded asteroid data
appeared ``smeared.''  In order to correct for the flux lost outside
the nominal 10-pixel-radius synthetic aperture, the nominal
\unit{11.88}{\micron} flux was increased by a factor of 13\%, derived
using the growth curve method \citep{Howell1989}. { To test the importance of this approximation we fit the data without this point, and also with this point with a 10\% and 20\% uncertainty, finding  at most a 3\% change in diameter and no change in $\eta$.}
See Table \ref{T_flux_VISIR} for final fluxes.





\subsection{Diameter and albedo results}
\label{sect:neatm}

We { used} the Near-Earth Asteroid Thermal Model \citep[NEATM,][]{neatm}%
\footnote{See also \citet{Delbo2002} for an introduction and
  overview.}  to fit the thermal fluxes reported in
Table~\ref{tab:mirsi} and \ref{T_flux_VISIR}. 
%
{
NEATM fluxes are calculated by integrating the Planck function
over the visible and illuminated part of the surface of a spherical
model asteroid.  The surface is assumed to be in
instantaneous thermal equilibrium with the absorbed sunlight, where
temperatures across the surface can be rescaled to match the spectral
energy distribution of the observed fluxes.  This temperature
rescaling is expressed in terms of a dimensionless ``beaming
parameter'' $\eta$ such that $T^4\propto \eta$.  
} 
{ Note that our observations do not spatially resolve \FS. Thus, the NEATM-derived diameter $D$ is that of a sphere with the combined cross-sectional area of the two components.  The (area-equivalent) component diameters $D_1$ and $D_2$ are related to $D$ via $D_1{}^2 + D_2{}^2 = D^2$.}

In order to determine the statistical uncertainty in the fit
parameters { of the} NEATM $D$, \pv, and $\eta$, we { performed} a Monte-Carlo analysis
analogous to that described by \citet{Mueller2007} in which 300 random
flux sets are generated, normally distributed about the measured data.
The median of the Monte-Carlo results is adopted as the nominal result
with asymmetric error bars encompassing the central 68.2\% of the
results \citep[see][for a more detailed discussion of this
  method]{Mueller2011}.

\begin{figure}[h]
\includegraphics[angle=90,width=\linewidth]{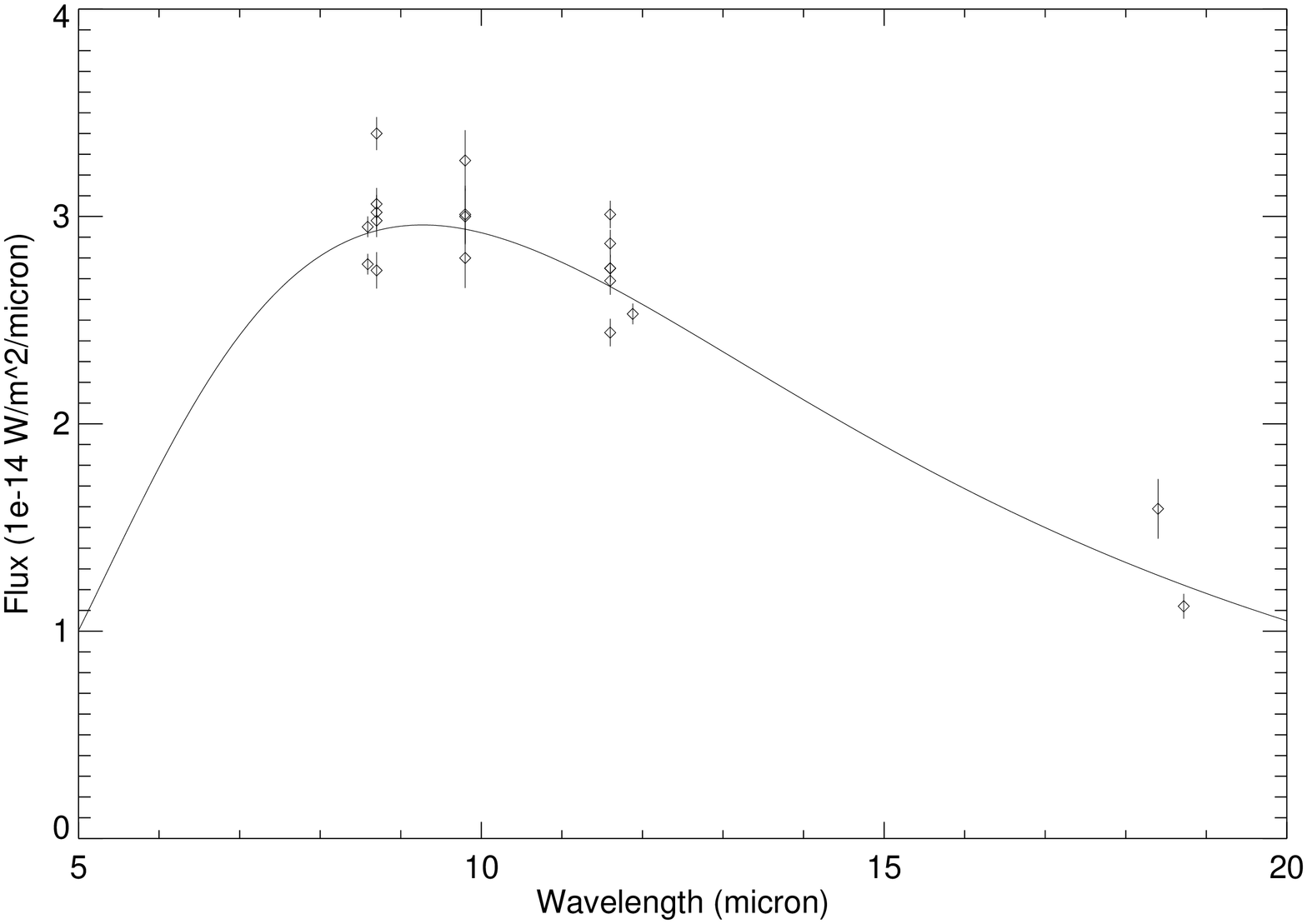}
\caption{Thermal-IR data of \FS\ (see Tables \ref{tab:mirsi} and \ref{T_flux_VISIR}) and  NEATM model curve using the adopted  parameters from Table \ref{tab:neatmresults}.
\label{fig:neatm}
} 
\end{figure}

This Monte-Carlo analysis was performed for each dataset (see Tables
\ref{tab:mirsi} and \ref{T_flux_VISIR}) and for their
combination---note that the observation geometry is essentially
identical between the two nights; the corresponding change in expected
flux is of the order of 1.6\%. 
Given the apparent discrepancy between the
Q-band fluxes measured during the two nights (see below), we repeat
this analysis for all datasets minus Q-band fluxes.  All results along
with Monte-Carlo uncertainties are given in Table
\ref{tab:neatmresults}.
The data are plotted in
Figure~\ref{fig:neatm} along with the adopted best-fit model curve.

The true uncertainties in $D$ and \pv\ are dominated by the systematic
uncertainty inherent in the NEATM, which is conventionally estimated
to be  15\% in $D$ and 30\% in \pv\ \citep{Harris2006}.  In the
(exceptional) case of \FS, the uncertainty in $H$ does not contribute
significantly to the error budget.  Our results
from the different datasets are in excellent mutual agreement; in
particular, the Q-band fluxes are largely inconsequential for our
purposes.  We adopt the average of the first six lines in Table
\ref{tab:neatmresults} as our final result, including the systematic
$D$ and \pv\ uncertainty; the adopted $\eta$ uncertainty is the
scatter between the six datasets:
{
$D=1.90\pm\unit{0.28}{\km}$, $\pv=0.039\pm0.012$, and $\eta=1.61\pm0.08$.
Using the known diameter ratio of $D_2/D_1 = 0.28^{+0.01}_{-0.02}$ \citep{Scheirich2009}, the corresponding (area-equivalent) component diameters are $D_1=1.83\pm\unit{0.28}{\km}$ and $D_2=0.51\pm\unit{0.08}{\km}$.}



Our results are in excellent agreement with those reported by
\citet{Mueller2011} based on Warm-Spitzer observations
($D=1.8^{+0.6}_{-0.5}\unit{{}}{\kilo\metre}$ and
$\pv=0.04^{+0.04}_{-0.02}$)
{
and with those reported by \citet{Wolters2011} ($D=1.71\pm\unit{0.07}{\km}$, $\pv=0.044\pm0.004$, and $\eta=1.15$).}
{ Note that \citeauthor{Wolters2011}'s observations took place at a significantly lower phase angle (11.7\degree) than ours (67.4\degree), hence the difference in best-fit $\eta$ values is consistent with expectations \citep{Delbo2007}.}
\citet{Mueller2011} did not constrain
$\eta$, but { assumed} a value.  
{ The good mutual diameter agreement of these three studies}
therefore provides independent support for the assumptions made by
\citet{Mueller2011}. 

\section{Visible-Near-IR reflectance and \FS's main-belt
  origin}\label{visnir}



\begin{figure}[h]
\includegraphics[angle=270,width=\linewidth]{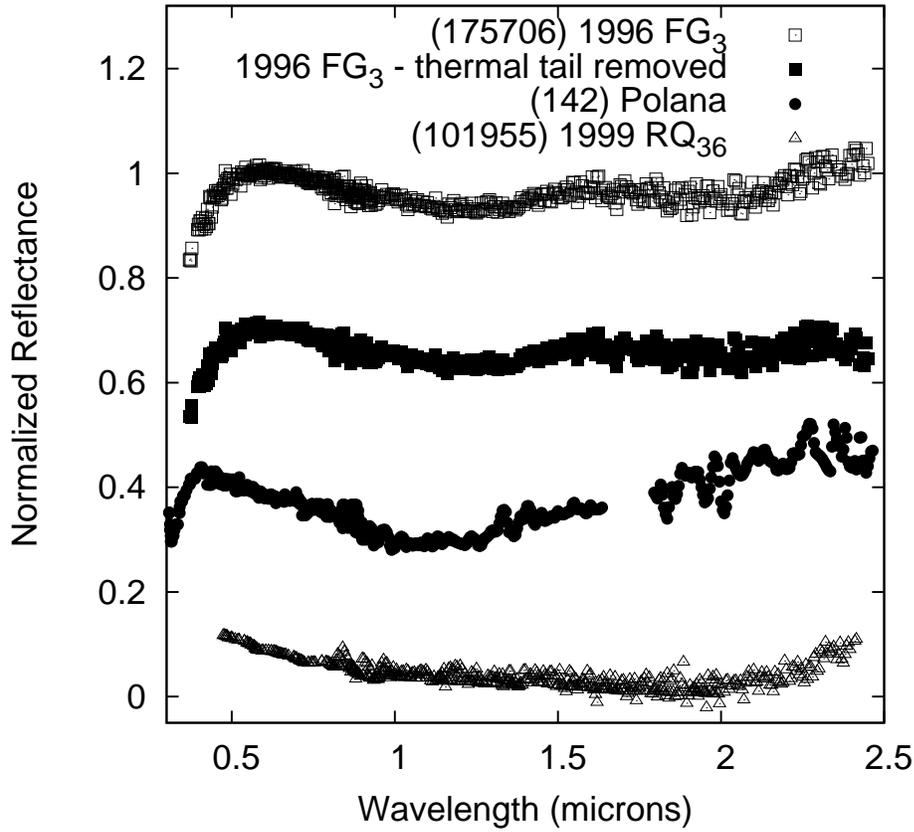}
\caption{Visible-near-IR spectral reflectance from 0.4--2.5 $\mu$m for
  1996 FG$_3$, 1996 FG$_3$ with the thermal tail removed, (142) Polana
  and 1999 RQ$_{36}$ (B. Clark et al. 2011 submitted). 
\label{spec}}
\end{figure}

As part of the MIT-UH-IRTF Joint Campaign for NEO Spectral
Reconnaissance the reflectance of \FS\ was measured { between 0.8
  and 2.5}
\micron. In Figure \ref{spec} the visible reflectance from the SMASS
survey is combined with the near-IR to provide reflectance from
0.37--2.5 \micron\ \citep{Binzel2004}. The spectrum presented was
measured on 30 March 2009.  A total integration time of 2880 seconds
was obtained during a one-hour interval beginning at 11:17 UT.  A
nearby solar analog star, Landolt 105-76, was observed at similar
airmass immediately following the asteroid measurements.




The 30 March 2009 spectrum of \FS\ { does not yield a unique class,
  rather the allowable classes of} C, Ch, or Xk type { was
  determined} by the online visible-near-IR taxonomy classifier based
on the DeMeo visible near-IR taxonomy
\citep{Demeo2009,DeLeon2011}. However, a B-type classification in the
DeMeo taxonomy demands a negative slope to 2.45 \micron, and in the
spectrum a thermal tail is evident starting $\sim$2.0 \micron\ { (see Fig. \ref{spec})}.  We fit
the NEATM to the thermal tail of the SpeX data using the method
described in \citet{Delbo2011}.  A large range in \pv--$\eta$
combinations fit the thermal tail well, though we use \pv=0.04,
$\eta\sim1.2$. The phase angle during the SpeX observations was
$\sim\unit{8}{\degree}$, i.e., much lower than during our mid-IR
photometric observations, hence such a low $\eta$ is consistent with
our expectations.

  The thermally-corrected spectrum was classified as a B-type by the
  online classification system \citep{Demeo2009}.  \citet{DeLeon2011}
  find that the 27 April 2009 spectrum also taken by the MIT-UH-IRTF
  campaign of lower SNR is classified as a B-type by the same online
  tool. However, their spectrum, taken at Telescopio Nazionale Galileo
  (TNG) on 9 January 2011, is labeled a Ch or Xk type.  We move
  forward with a B-type classification, though naming a taxonomic type
  is less indicative of composition than an analysis of spectral
  features. { Therefore we provide a quantitative comparison and a
    qualitative interpretation of inferred spectral features.}



  { The $\approx$ 50~km Main Belt asteroid (142) Polana is the largest
    member of a family proposed to have been the main-belt origin of
    B-type NEO 1999 RQ$_{36}$ \citep{Campins2010}, due in large part
    to its B-type spectrum.  To quantify the similarity between these
    spectra we follow the ${\chi}^2$ formulation used by
    \citet{deLeon2010} and \citet{Campins2010} between 0.475 and
    1.925
    \micron\ with spacing of 0.0025 \micron.  \citet{Campins2010}
    found a $\chi^2 = $ 4.91 when comparing 1999 RQ$_{36}$ with (142)
    Polana. Using a similar, but not exact, wavelength range we find a
    value $\chi^2 = $ 4.83 for these two objects. \FS\ compared to
    1999 RQ$_{36}$ and (142) Polana yield values of $\chi^2 = $ 1.09
    and 1.86, showing a strong similarity to both. The closest fit
    found by \citet{Campins2010} for 1999 RQ$_{36}$ and 27 main belt
    B-type asteroids was $\chi^2 = 1.33$, so its likeness to another
    B-type such as \FS\ is naturally greater than that to the other
    main belt bodies in that study. Similarly, the match between
    \FS\ and (142) Polana is also closer than for 1999 RQ$_{36}$ and
    (142) Polana. However, when comparing spectra via a $\chi^2$
    routine, features representing mineralogical differences can be
    marginalized in favor of bulk shape or slope.

}






There are subtle but clearly revealed differences between the three
asteroids (Figure \ref{spec}) that may allow differing mineralogical
interpretations.  The features near 1.2 and 2.0 \micron\ for \FS\ are
consistent with the presence of some olivine and some pyroxene, where
the shallow depth of the features could be due to opaques. Note,
however, that \citet{DeLeon2011} rejects these two features due to the
shape and location of the first band. This is an important
disagreement, as our interpretation imply silicates, whereas
\citet{DeLeon2011} does not.  The deeper 1.2 \micron\ feature for
Polana, with no feature at 2.0 \micron\ is consistent with the
presence of olivine (and perhaps opaques).  Having spectral hints of
olivine and pyroxene is possibly consistent with the spectral
properties of ureilite meteorites \citep{Cloutis2010}, whereas the
interpretation of \citet{DeLeon2011} is more suggestive of
carbonaceous chondrites (they suggest CM2).  However, 1999 RQ$_{36}$
is itself distinct from the others for its lack of 1 and 2 \micron\
spectral features, which is more typical of carbonaceous chondrites.

  \citet{DeLeon2011} also made comparisons with meteorite spectra from
  the RELAB databse. Their TNG spectra was best fit by a combination
  of CM2 carbonaceous chondrites, and L4 and H5 ordinary
  chondrites. Their preferred match for the high-SNR MIT data was with
  a weakly shocked H4 ordinary chondrite.





\section{Dynamical history and \FS's main-belt origin}\label{dyn}
The chaotic nature of NEA orbital evolution makes it impossible to
know the exact history for any given body. However, detailed orbital
evolution modeling by \citet{Bottke2002} 
allows probabilistic estimates for Main Belt source regions to be
made. 
The \citeauthor{Bottke2002}\ work finds that the current orbit of
\FS\ supports a { 92\%} likelihood that it became an NEA after escaping
the { inner} Main Belt via the $\nu_{6}$ secular resonance\footnote{ This
  probability was determined using the 3-source NEO model, where 3
  Main Belt sources are modeled as the parent populations for NEOs.
  This is the technique used for 1999 RQ$_{36}$ by
  \citet{Campins2010}.}.

The $\nu_{6}$ marks the inner limit of the main asteroid belt and thus
asteroids only reach this resonance by migrating inward from orbits
with a larger semi-major axis ($a$ $>$ 2.15 AU).  Similarly, this
region is bounded at large semi-major axis by the 3:1 mean motion
resonance (MMR) with Jupiter located at 2.5 AU.  Thus, escape by the
$\nu_{6}$ resonance requires an initial semi-major axis range of 2.15
$< a <$ 2.5 AU, { and inward migration}. Escape by the $\nu_{6}$ largely retains the
inclination of the orbit so that the current 1.99$^\circ$ inclination
is likely near its original inclination in the Main Belt. 


A similar analysis has been done for the NEA 1999 RQ$_{36}$
where its orbital, physical and rotation properties were used to infer
a parent family in the inner main belt. Its match with the (142)
Polana family is due in large part to its classification as a B-type
asteroid, and similarities in their visible/near-IR
spectra. Dynamically, 1999 RQ$_{36}$ was found to have a 95\%
probability of coming from the inner belt via the $\nu_6$ resonance,
and has a low inclination of $i \sim 6.04^{\circ}$. 1999
RQ$_{36}$ is in retrograde rotation, and also has a ``top-shape''
which has been found frequently among binary NEAs
\citep{Nolan2007,Harris2009}. 

Despite uncertainties in the { Polana} family's age, small members {
  (H~$>$~18.5)} have already reached the $\nu_6$ resonance by the
Yarkovsky effect \citep{Campins2010}. The retrograde rotation of 1999
RQ$_{36}$ also indicates that its Yarkovsky drift is inward, allowing
it to drift from the (142) Polana family towards the $\nu_6$
resonance.
 
{ \FS\, however, has H=$17.76$ and thus should not have had time to
  reach the $\nu_6$ resonance by the Yarkovsky effect according to
  this estimate of the Polana family extent \citep{Campins2010} (Note
  also that no definitive age is known for Polana due to its member's
  overlap with the Nysa family). However, it is possible for objects
  the size of \FS\ to reach the overlapping Jupiter 7:2 and the Mars
  5:9 resonances located at $a$=2.2559-2.2569~AU and $a$=2.2542-2.2550~AU
  \citep{MorbidelliNesvorny1999,Bottke2007}.
  This region of overlapping resonances is discussed in the
  Supplementary Material of \citet{Bottke2007}, but its efficiency as
  a source is uncertain for objects of \unit{2}{\km} such as
  \FS. Figure S6 of \citet{Bottke2007} shows roughly equal numbers of
  \unit{5}{\km} bodies jumping the resonance or getting trapped and
  excited.

  Smaller objects will have higher Yarkovsky drift rates, likely
  lowering the fraction that get excited onto planet-crossing orbits.
  We tested the efficiency of resonance-crossing for $\sim
  \unit{2}{\km}$ bodies drifting at their maximum Yarkovsky drift
  rates, which provides a lower limit. We used a modified version of
  {\tt swift\_rmvs} that applied a constant change in semi-major axis
  of -10$^{-4}$~AU Myr$^{-1}$ as estimated from \citet{Bottke2006} as
  an upper limit for its Yarkovsky drift rate at this size. We tested
  a set of 155 massless test particles initially located at $a
  \approx$2.26~AU. The particles had initial $e$ and $i$ values
  similar to those of Polana, and 21 of the 155 particles had their
  eccentricities excited to Mars-crossing levels while crossing the
  resonance, and evolving onto Mars-crossing orbits. From here their
  evolution onto NEO orbits will be relatively rapid and leave them
  virtually indistinguishable from asteroids entering NEO-space via
  the $\nu_6$ resonance.

  The 21 particles excited out of the inner Main belt all had initial
  eccentricities above 0.15, pointing to the resonance's increasing
  effectiveness for higher $e$. The Polana family itself covers a
  range from $e$=0.13--0.17, even though (142) Polana itself has an
  $e$=0.136. Even with the conservative assumption of the
  maximum Yarkovsky drift rate for a $D$=2~km asteroid, approximately
  10~\% of the of family members this size would exit the inner Main
  Belt by the overlapping J7:2 and M5:9 resonances. Therefore,
  \FS\ could still have an origin with the Polana family as it fits
  the other major criteria used by \citet{Campins2010} to locate (142)
  Polana as a possible parent family of 1999 RQ$_{36}$: a B-type
  asteroid, a likely inner-Main Belt origin, low inclination,
  retrograde rotation. Making this study more general, to estimate the
  number of Polana members of this size in the NEO population, is
  challenging due to the uncertainties surrounding the exact size and age of
  the Polana family.



}


\section{Results and Conclusions}

The findings of these observations and dynamical studies are;
\begin{enumerate}
\item{} \FS\ has a diameter of $1.90 \pm \unit{0.28}{\km}$ and a geometric albedo of $\pv=0.039 \pm 0.012$,

\item{} { Available data do not uniquely place \FS\ into a single taxonomic class; the best data (highest SNR) with a thermal flux correction applied, are most consistent with the B-class,}
\item{} \FS\ is likely ({ 92\%}) to have come from the inner Main Belt via
  the $\nu_6$ resonance, { though it could have escaped by the overlapping J7:2/M5:9 resonances located at 2.255~AU.}
\end{enumerate}

Together, these results provide a much better picture of the origin of
\FS\, and raise some tantalizing questions about its relation to the NEA
1999 RQ$_{36}$, another ideal space mission target and primary
target of NASA's OSIRIS-REx sample return mission
\citep{Lauretta2010}.


\citet{Delbo2011} studied the thermal properties of binary NEAs which
have been observed in the thermal-IR, finding that statistically
elevated $\eta$ values, used as a proxy for thermal inertia, pointed
to cooler surfaces than non-binary NEAs. This was interpreted in terms
of regolith loss during the binary-formation process. Here, \FS\ was
fit with an $\eta$ of $1.61\pm0.08$.  The large phase angle
$\sim\unit{68}{\degree}$ makes this less constraining, but the values
are elevated compared to the sample presented in
\citet{Delbo2011}. 
{ \citet{Wolters2011} obtained $\eta\sim1.15$ at a phase angle of \unit{11.7}{\degree}.  
Taken together, the available thermal-IR data should enable a robust determination
 of the thermal inertia; that will be discussed, however, in a forthcoming paper (Mueller et al., 2012, in preparation).
In comparison, the thermal inertia of 1999~RQ$_{36}$ has been estimated to be
around \unit{600}{\tiu} \citep{Emery2010}. Based on the correlation
between thermal inertia and object size found by \citet{Delbo2007}, and
using $D\sim\unit{0.58}{\km}$ \citep{Nolan2007}, a thermal inertia around
\unit{400}{\tiu} would be expected, somewhat lower than measured.
While 1999~RQ$_{36}$ is not known to be a binary, its radar shape, oblate
with a possible equatorial ridge, is suggestive of YORP spinup and
reshaping. Like in the case of YORP binaries, this could lead to an
elevated thermal inertia such as that given by \citet{Emery2010}.
%
%
}

\FS\ and 1999 RQ$_{36}$ share many physical properties and possibly a
similar dynamical history. As shown in Section \ref{visnir}, there is
some uncertainty about the precise taxonomy for \FS, though its
spectra is qualitatively similar, with a  negative slope from
0.5-2.0 \micron, to that of 1999 RQ$_{36}$. The features for \FS\ at
1.2 and 2.0 \micron\ are an important distinction between \FS\ and
1999 RQ$_{36}$. If these features are indicative of silicates then
\FS\ may have more similarities with ureilites than with carbonaceous
chondrites.  In contrast, the featureless spectra of 1999 RQ$_{36}$ is
suggestive of properties similar to carbonaceous chondrites, with CM
typically cited as close matches for ``B-types'' depending on the
level of thermal processing \cite{DeLeon2011,Clark2010,Hiroi1996}.

It is important to note that \citet{Jenn2009} obtained a low
signal-to-noise visible spectrum of asteroid 2008 TC$_3$ prior to its
Earth impact and subsequent recovery as the Almahata Sitta meteorite.
While their initial interpretation of this spectrum suggested F- or
B-type classification in the system of \citet{Tholen1989}, that
classification is not confirmed with the addition of near-infrared
spectrum of a Almahata Sitta meteorite presented by \citet{Jenn2010}
and \citet{Hiroi2010}. While a composite average spectrum of different
stones of Almahata Sitta reproduces the 2008 TC$_3$ visible spectrum,
none of the measured lab spectra matches the B-class (or any asteroid
class defined over the full spectral range defined by
\citet{Demeo2009}).

We also note that the F-class itself diverges (fails to exist) for
objects whose spectra are extended from the visible wavelengths to the
near-infrared.  As discussed by \citet{Demeo2009}, objects that are
spectrally similar over visible wavelengths and satisfy the
\citet{Tholen1989} definition of the ``F-class" diverge from one
another as their spectra are extended into the near-infrared.  Thus at
present, the visible plus near-infrared spectrum of the Almahata Sitta
meteorite spectrum fits into no asteroid taxonomy class and closely
resembles no other currently measured spectrum of any asteroid.



\citet{Jenn2010} also pursued a parent source family for 2008 TC$_3$
similar to the dynamical work of \citet{Campins2010} for 1999
RQ$_{36}$.  The most probable source region was found to be the inner
main asteroid belt via the $\nu_6$ resonance ($\sim$80\% probability):
the same source region as \FS\ and 1999 RQ$_{36}$.  \citet{Jenn2010}
speculated the (142) Polana family as a source for 2008
TC$_3$. However the jumbled asteroid-meteorite link for the spectrum
of 2008 TC$_3$ renders any further connection beyond the scope of this
work.


Thus the implication of this study are that two prime space-mission
targets, 1999 RQ$_{36}$ and \FS\ share similar taxonomies, spectra and
possibly even rotation history (YORP-spinup). { Linking them to the
  same parent family in the inner-main belt, the Polana family, is
  highly speculative as \FS\ would have had to escape by a minor
  resonance rather than $\nu_6$. However, it cannot be ruled out
  entirely as we have shown that the J7:2/M5:9 resonances can push
  asteroids these size onto Mars-crossing orbits and on the path to
  near-Earth orbits.}

\section*{Acknowledgments}
Thanks to Tom Burbine for a helpful discussion, and David Nesvorn{\'y} for
his version of {\tt swift\_rmvs3}.  MM and RPB are Visiting Astronomers
at the Infrared Telescope Facility, which is operated by the
University of Hawaii under Cooperative Agreement no. NNX-08AE38A with
the National Aeronautics and Space Administration, Science Mission
Directorate, Planetary Astronomy Program.  The great work of the IRTF
team, and particularly of the TOs, is gratefully appreciated. Part of
the data utilized in this publication were obtained and made available
by the The MIT-UH-IRTF Joint Campaign for NEO Reconnaissance.  The MIT
component of this work is supported by the National Science Foundation
under Grant No. 0506716.



\newpage

\newpage

\begin{figure}[h]
\includegraphics[width=\linewidth]{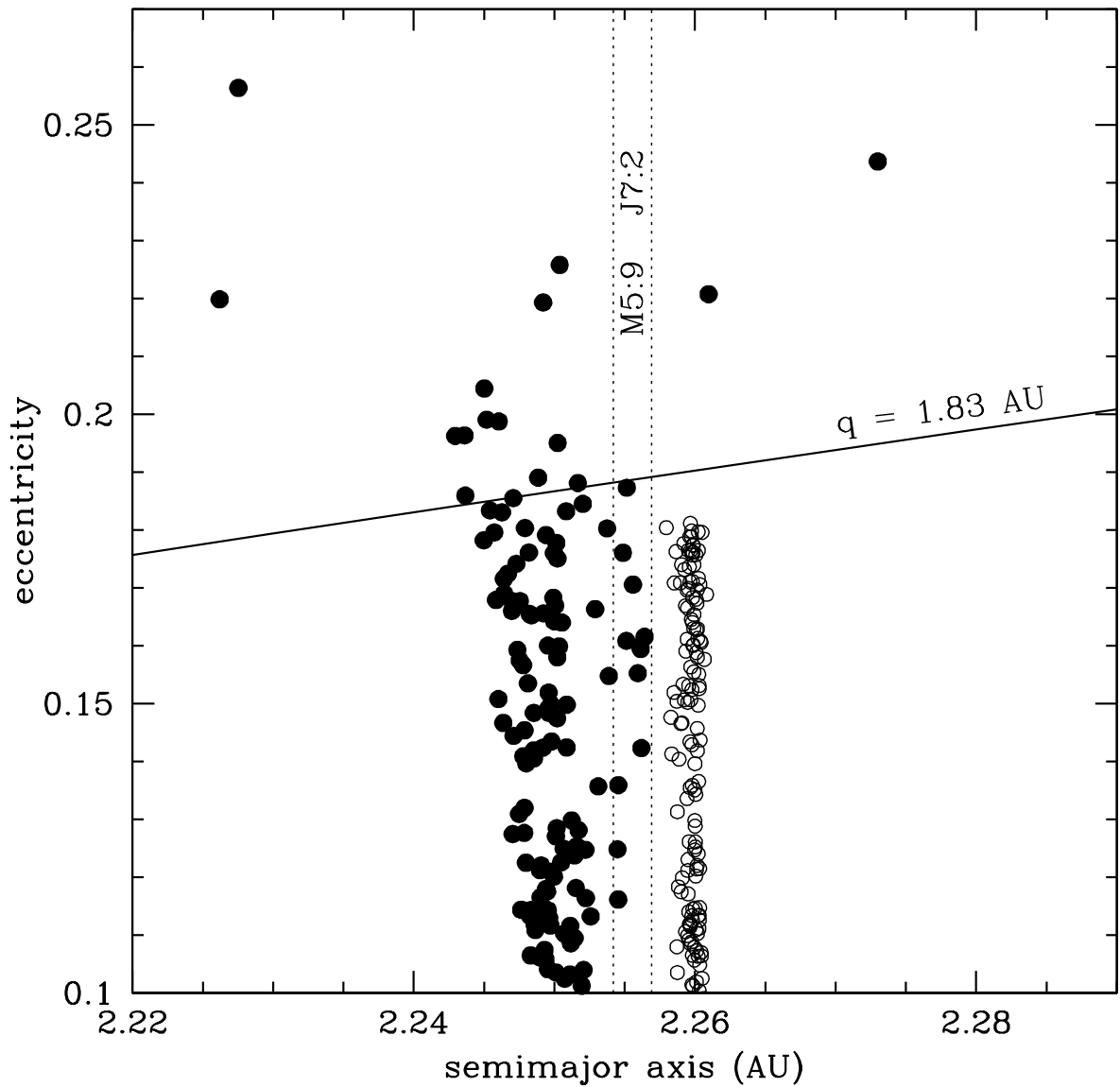}
\caption{The evolution of massless test particles evolving while a
  constant change in semimajor axis of $10^{-4}$~AU/MYR is
  applied. The plot shows the eccentricity and semimajor axis for each
  of the 155 particles at the start of the integration (open symbol)
  and at the end of the 100~Myr integration (closed symbols).  The
  vertical dashed lines are the location of the Jupiter 7:2 MMR and
  the Mars 5:9 MMR, while the horizontal dashed line represents a
  Mars-crossing orbit at $q=1.83$~AU
\label{KOMR}}
\end{figure}


\newpage

\begin{table}[tb]
\caption{
\label{tab:mirsi}
MIRSI observations of \FS\ on 2009 May 1 UT.
\FS\ was at a heliocentric distance of
$r=\unit{1.057}{AU}$, observer-centric distance of
$\Delta=\unit{0.1567}{AU}$, and phase angle of
$\alpha=\unit{67.4}{\degree}$; all values were constant during our
observations to within the last quoted digit.  }

\small
\begin{tabular}{lr@{.}lcc}
\hline
\hline
UT  & \multicolumn{2}{c}{Wavelength} & Flux & Flux \\
    & \multicolumn{2}{c}{\micron}& Jy   & $10^{-14}$\watt\per\meter\squared\per\micron \\
\hline
05:51 & 11&6 & $1.23\pm0.03$ & $2.75\pm0.07$ \\
05:53 & 11&6 & $1.29\pm0.03$ & $2.87\pm0.07$ \\
05:54 & 11&6 & $1.35\pm0.03$ & $3.01\pm0.07$ \\
05:59 &  8&7 & $0.86\pm0.02$ & $3.40\pm0.08$ \\
06:01 &  8&7 & $0.77\pm0.02$ & $3.06\pm0.08$ \\
06:10 & 18&4 & $1.80\pm0.16$ & $1.59\pm0.14$ \\
06:17 &  9&8 & $0.96\pm0.04$ & $3.00\pm0.13$ \\
06:20 &  9&8 & $1.05\pm0.05$ & $3.27\pm0.15$ \\
06:23 &  9&8 & $0.96\pm0.04$ & $3.01\pm0.14$ \\
06:25 & 11&6 & $1.23\pm0.03$ & $2.75\pm0.06$ \\
06:27 & 11&6 & $1.21\pm0.03$ & $2.69\pm0.07$ \\
06:30 &  8&7 & $0.75\pm0.02$ & $2.98\pm0.08$ \\
06:33 &  8&7 & $0.69\pm0.02$ & $2.74\pm0.09$ \\
06:36 &  8&7 & $0.76\pm0.02$ & $3.02\pm0.08$ \\
06:43 & 18&4 & $1.80\pm0.16$ & $1.59\pm0.14$ \\
06:48 & 11&6 & $1.10\pm0.03$ & $2.44\pm0.07$ \\
06:51 &  9&8 & $0.90\pm0.05$ & $2.80\pm0.15$ \\
\hline
\end{tabular}
\end{table}

\newpage

\begin{table}[h!] \small
\begin{tabular}{cr@{.}lcc}
    \hline
\hline
UT  &  \multicolumn{2}{c}{Wavelength} & Flux & Flux \\
    & \multicolumn{2}{c}{\micron}& Jy   & $10^{-14}$\watt\per\meter\squared\per\micron \\
     \hline
  00:56 & 8&59     & 0.68$\pm$0.01 & 2.95$\pm$0.05 \\
  01:14 & 8&59     & 0.73$\pm$0.01 & 2.77$\pm$0.05 \\
  00:28 & 11&88    & 1.19$\pm$0.02 & 2.53$\pm$0.05 \\
  01:05 & 18&72    & 1.31$\pm$0.07 & 1.12$\pm$0.06 \\
\hline
\end{tabular}
\caption{VISIR observations of \FS\ on 2 May 2009 UT.  The target was at an observer-centric distance 
  of 0.156 AU, a heliocentric distance of 1.053 AU, and a
  phase angle of 69.1$^\circ$. Note: the \unit{11.88}{\micron} flux quoted here
includes the 13\% smearing correction discussed in the text.
}
\label{T_flux_VISIR}
\end{table}

\newpage

\begin{table}[tb]
\caption{NEATM fit results for \FS. Each dataset is analyzed with and without Q-band data. Adopted final results are given in the last line. The uncertainties in the first six lines are purely statistical, whereas the last line includes systematics.
\label{tab:neatmresults}
}
\begin{tabular}{r c c c}
Data set & $D (\kilo\metre)$ & \pv & $\eta$ \\
\hline
MIRSI+Q & $1.95^{+0.06}_{-0.07}$ & $0.036^{+0.003}_{-0.002}$ & $1.67^{+0.08}_{-0.13}$ \\
MIRSI-Q & $1.90^{+0.07}_{-0.06}$ & $0.039^{+0.003}_{-0.003}$ & $1.59^{+0.10}_{-0.10}$ \\
VISIR+Q & $1.81^{+0.07}_{-0.07}$ & $0.043^{+0.004}_{-0.003}$ & $1.51^{+0.10}_{-0.09}$ \\
VISIR-Q & $1.84^{+0.11}_{-0.06}$ & $0.041^{+0.003}_{-0.005}$ & $1.55^{+0.14}_{-0.06}$ \\
Both+Q  & $1.94^{+0.03}_{-0.06}$ & $0.037^{+0.003}_{-0.001}$ & $1.67^{+0.03}_{-0.08}$ \\
Both-Q  & $1.96^{+0.05}_{-0.06}$ & $0.036^{+0.003}_{-0.002}$ & $1.69^{+0.06}_{-0.09}$ \\
\hline
Adopted & $1.90\pm0.28$ & $0.039\pm0.012$ & $1.61\pm0.08$ \\
\end{tabular}
\end{table}

\end{document}